\documentclass{llncs}

\usepackage[english]{babel}

\usepackage{amsmath}
\usepackage{graphicx}
\usepackage[colorlinks=true, allcolors=blue]{hyperref}
\usepackage{xspace}

\newcommand{\registered}[0]{\textsuperscript{\textregistered}\xspace}
\newcommand{\trademark}[0]{\texttrademark\xspace}

\usepackage{algorithm}
\usepackage{algorithmicx}
\usepackage{algpseudocode}
\usepackage{subcaption}

\usepackage{xcolor}
\usepackage[final]{changes}
 
\definechangesauthor[name=arXiv-feedback, color=olive]{F}
\definechangesauthor[name=Review1, color=blue]{R1}
\definechangesauthor[name=Review2, color=red]{R2}
\definechangesauthor[name=Review3, color=brown]{R3}
\definechangesauthor[name=Review4, color=green]{R4}
\definechangesauthor[name=Adam, color=orange]{AT}

\title{Detrimental task execution patterns in mainstream OpenMP\registered runtimes}
\titlerunning{Task execution patterns}

\author{
  Adam S.~Tuft\inst{1}~\orcidID{0009-0001-0251-7041} 
  \and\\
  Tobias Weinzierl\inst{1}~\orcidID{0000-0002-6208-1841}
  \and \\
  Michael Klemm\inst{2,3}~\orcidID{0000-0002-8634-4634} 
}

\authorrunning{M.~Klemm and A.~Tuft and T.~Weinzierl}

\institute{
  Department of Computer Science, Durham University, Durham, United
  Kingdom \email{\{adam.s.tuft,tobias.weinzierl\}@durham.ac.uk}
  \and
  Advanced Micro Devices GmbH, Dornach/Munich, Germany
  \and
  OpenMP Architecture Review Board, Beaverton, OR, USA
  \email{michael.klemm@\{amd.com,openmp.org\}}
}

\begin{document}

\maketitle

\begin{abstract}
  The OpenMP\registered API offers both task-based and data-parallel concepts to scientific
computing.
While it provides descriptive and prescriptive annotations, it is in many places
deliberately unspecific how to implement its annotations.
As the predominant OpenMP implementations share  design rationales,
they introduce ``quasi-standards'' how certain annotations behave.
By means of a task-based astrophysical simulation code, we
highlight situations where this ``quasi-standard'' reference
behaviour introduces performance flaws.
Therefore, we propose prescriptive clauses to constrain the OpenMP
implementations.
Simulated task traces uncover the clauses' potential, while a
discussion of their realization highlights that they would manifest in
rather incremental changes to any OpenMP runtime supporting task priorities.


\end{abstract}



\pagenumbering{arabic}
\pagestyle{plain}

\section{Introduction}
\label{section:introduction}

%
%
With the advent of hundreds of cores on a contemporary computer chip in data
centres, classic data parallelism reaches scalability limits.
Even if we decompose algorithms into sequences of highly parallel steps, we will eventually
fail to exploit the available parallelism of a machine.
Task-based programming promises to ride to our rescue.
From a programmer's point of view, it imitates object-orientation's
success stories.
Rather than reading an algorithm as a sequence of steps where each step exploits
parallel capabilities over a large data set, we decompose
an algorithm into many small ``mini-algorithms'' over well-defined
data sets, i.e., all the data they actually read and write.
The tasks can then spawn further child tasks or have inter-task dependencies.

A task logically encapsulates data plus operations on these data.
While programming with tasks might reflect programming's best practices,
the HPC selling point behind tasks results from the fact that they help us to
expose unprecedented scheduling freedom:
task dependencies can often replace synchronization in-between algorithmic steps.
Tasks allow us to write code with a high theoretical concurrency level.

%
%
The OpenMP\registered API~\cite{OARB21} has offered task directives since version 3.0, which was
released in 2008.  Since then, the task features
have been refined and extended to provide a state-of-the-art task-parallel
programming interface.  
The OpenMP specification does
not require a specific execution mechanism or implementation strategy for
OpenMP tasks.
It is deliberately unspecific in several places, and therefore offers a
certain degree of freedom to its implementations.
\replaced[id=R1]{A mainstream}{The most commonly used} OpenMP 
\replaced[id=R1]{implementation is provided with}{task implementations are the
ones of} the Clang/LLVM compiler and \added[id=R1]{all of our
experimental data stem from this ecosystem.}\replaced[id=F]{
 Though \replaced[id=R1]{there are alternative popular implementations such as
 GNU's OpenMP runtime, all}{completely different implementations, both}
 share similar design rationale }{
the GNU Compiler Collection.
Both employ similar implementation ideas and rationale}
\cite{KlCo21}.
\replaced[id=R1]{Whenever they}{Both therefore} yield similar execution
patterns for a given task graph\added[id=R1]{, our observations and concepts
apply. They are generalizable}.
\deleted[id=F]{They exploit implementation freedom in similar ways.}

%
%
In this paper, we study a numerical astrophysics code based upon
adaptive Cartesian meshes \cite{Zhang:2024:ExaGRyPE} which heavily relies upon tasking
\cite{Charrier:2020:EnclaveTasking,Li:2022:TaskFusion}.
We use it to highlight where the execution patterns from predominant
OpenMP implementations are detrimental to the code's performance.
Through an artificial scheduling model, i.e., a task schedule simulator, we are
able to quantify what better performance alternative schedules might be able to
deliver.

%
%
Once we have introduced our demonstrator and the simulator
(Sect.~\ref{section:demonstrator}),
we discuss, per runtime flaw, the degree to which it is a result
of the OpenMP specification or arises from implementation decisions
(Sect.~\ref{section:patterns}).
In Sect.~\ref{section:realisation}, we propose extensions to the OpenMP tasking API which would
allow an application to manipulate the task execution pattern and
hence to run faster.
Challenging the well-intended rationale of some OpenMP
implementations, our work stands in the tradition of a transition from a descriptive to a prescriptive
parallelization model.
We conclude in Sect.~\ref{section:conclusion} that programmers should, if they
want, have a stronger say in how a task graph is actually mapped onto a task schedule.

\section{A stationary black hole simulation analysed with
Otter}
\label{section:demonstrator}

We illustrate all OpenMP behaviour by means of a demonstrator from our
astrophysical simulation suite ExaGRyPE~\cite{Zhang:2024:ExaGRyPE}.
It simulates a single, stationary black hole that is modelled via a first-order
CCZ4 formulation \cite{Dumbser:2018:CCZ4}.
Various numerical building blocks feed into this simulation, ranging from
higher-order methods, adaptive mesh refinement, Sommerfeld boundary conditions,
to tracer particles that allow us to evaluate global integrals over
submanifolds.

This is an artificial yet numerically challenging
setup~\cite{Zhang:2024:ExaGRyPE}.
To tackle the scenario, we need to simulate a large computational domain
over a long time span exploiting all compute capabilities of the machine
efficiently.


\subsection{ExaHyPE's code architecture}

ExaGRyPE is a suite of solvers built on top of ExaHyPE
\cite{Reinarz:2019:ExaHyPE} and its meshing framework Peano.
Peano's adaptive mesh refinement (AMR) is mandatory to zoom into the area
around the black hole.
The arising adaptive mesh is static, i.e., it does not change over time.
We use plain domain decomposition along the Peano space-filling curve to
decompose the mesh for multiple processes using MPI.
The same non-overlapping domain decomposition is then used once more to split up
the rank-local domain and to distribute the arising chunks of the domain among
the available threads.
Per rank, this yields a classic fork-join parallelism.
Within Peano, we map it onto an OpenMP \texttt{taskloop}.
Each task traverses one subdomain on the rank, triggers all the simulation computations, and eventually synchronizes the
subdomain-local data (for example, halos) with other tasks and ranks.
The number of these traversal tasks is typically relatively small, as the data
synchronization towards the end quickly eats up all efficiency gains if we make
the subdomains too small.

\begin{algorithm}[htb]
  \scriptsize
  \begin{algorithmic}[1]
 \Function{traversal}{\ldots}
 \State \#pragma omp taskloop nogroup untied
 \For{(int subdomain = 0; subdomain $<$ K; subdomain++)}
   \For{(int cell\_in\_subdomain = 0; ...)}
     \Comment Traverse local subdomain
   \If{...}
     \Comment Only some cells define (produce) enclave tasks
     \While{database does not contain outcome yet}
       \State \#pragma omp taskyield
       \Comment Enclave outcomes from previous traversal
     \EndWhile
     \State \ldots
     \State \#pragma omp task
       \Comment Spawn enclave task
     \State \{
     \State \phantom{xxx} \ldots
       \Comment Actual work (*)
     \State \phantom{xxx} \ldots
       \Comment Dump outcome into database
     \State \}
   \Else
     \State \ldots 
       \Comment Process actual work immediately (*)
   \EndIf
   \State \ldots
   \EndFor
 \EndFor
 \State \#pragma omp taskwait
   \Comment Only wait for traversal tasks, not enclaves
 \EndFunction
\end{algorithmic}

   \caption{
    Pseudo code of the main traversal routine in ExaHyPE.
    \added[id=R2]{
      Algorithmic steps marked with an asterix host a
      \texttt{parallel for} loop.
    }
    \label{algorithm:travesal}
  }
\end{algorithm}

On top of the geometric data decomposition, we identify mesh cells or patches
which are free of side effects \cite{Li:2022:TaskFusion}, i.e., do not contribute
towards global quantities, and are not urgent in the sense that
they feed into MPI data exchange or AMR inter-resolution transfer operators.
The remaining cells or patches can be spawned as separate tasks with no further
in-dependencies or additional internal synchronization points.
They form
enclave tasks \cite{Charrier:2020:EnclaveTasking}, which can, without a
knock-on effect on MPI and the global simulation state, be executed at a later
point, i.e., even after the actual traversal.
The mesh traversal tasks therefore act as producers and consumers
of tasks, as enclave tasks are spawned in one mesh traversal and contribute
towards the solution in the subsequent mesh sweep
(Alg.~\ref{algorithm:travesal}).

Several compute steps both within the traversal and within the enclave tasks
exhibit further internal concurrency.
This manifests as loop parallelism
resulting from nested loops over different Finite Volumes or
sample points of the solution that evaluate the differential equations,
interpolate from one mesh resolution to another, or couple different numerical
schemes.


Despite the static nature of the setup and the adaptive mesh, the compute cost
per cell might change in each and every time step, as we
employ non-linear equation \replaced[id=R2]{solvers}{solves} with a dynamic
termination criterion per cell
\cite{Dumbser:2018:CCZ4,Reinarz:2019:ExaHyPE}.
With the adaptive mesh refinement, the projections along AMR boundaries are
expensive and make cells adjacent to AMR more costly than others.
The cost per AMR transition depends on its orientation, i.e., whether the data
layout favours a direction or not.
Finally, some of our ExaGRyPE solvers switch to a
subgrid model around the black hole.
This yields a very high load for some mesh cells compared to others.
W.l.g.~we use the compute cell count as the cost metric and therefore 
renounce the construction of a bespoke geometric load
balancing.
We accept that the outermost
fork-join parallelism due to traversal tasks is ill-balanced.

\subsection{Otter tracing}

\begin{figure}[t]
  \begin{center}
    \includegraphics[width=0.98\textwidth]{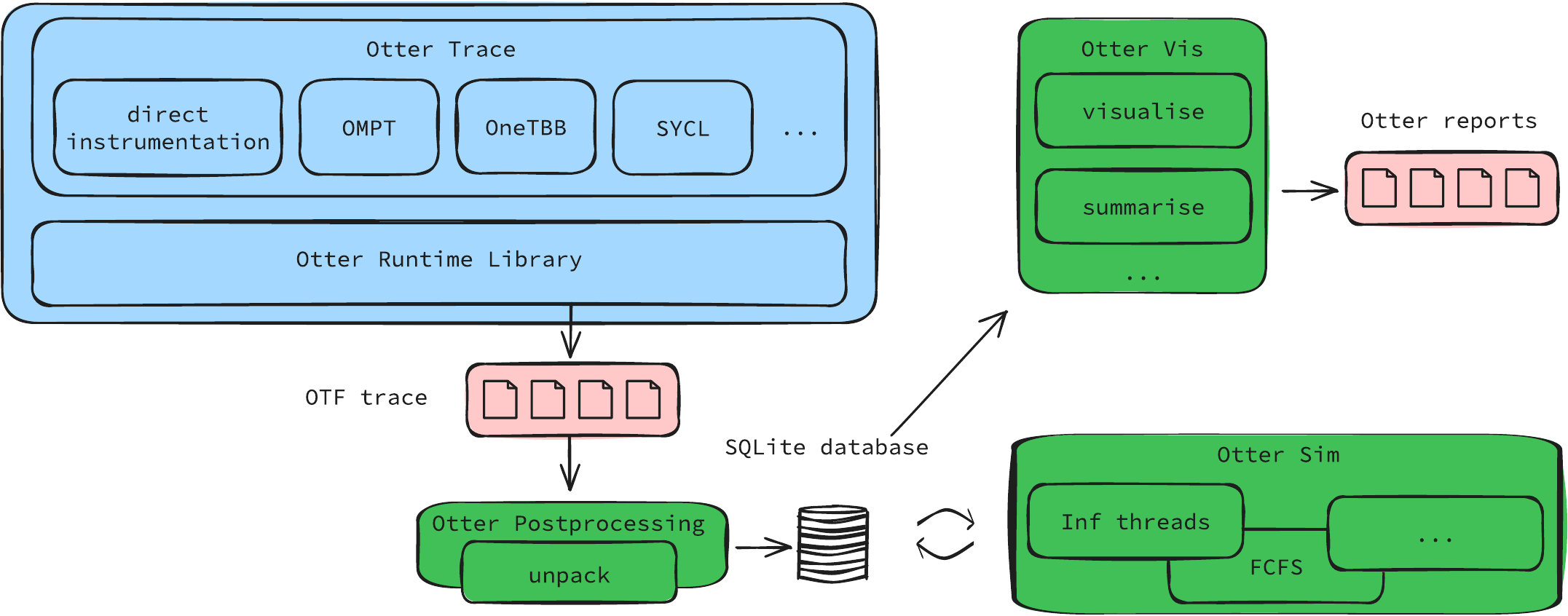}
  \end{center}    
  \vspace{-0.4cm} 
  \caption{
    The Otter tool suite and Otter's trace-simulate-postprocess workflow.
    \label{fig:otter-overview}
  }
\end{figure}

To study task execution patterns, we rely on a tool suite called
Otter (Fig. \ref{fig:otter-overview}). 
Otter offers a macro API for annotating serial or
partially parallelized code to highlight where tasks and loop parallelism could
be introduced theoretically.
It also can record existing OpenMP tasking through OMPT bindings. 
With both types of information, we run our simulation and let
\emph{Otter Trace} record the logical (hypothetical) task graph.
This trace includes timing data, too.
All trace data ends up in a modified OTF2 database \cite{Knupfer:2012:OTF2}.

Once postprocessed, scheduling simulators within the \emph{Otter
Sim} package allow us to re-play the recorded logical task graph within various
idealized schedulers assuming infinite thread counts, infinite task
queues serving all threads FCFS, different NUMA topologies, and so forth.
\emph{Otter Sim} can always (retrospectively) identify the critical path of a
code as it has access to the whole execution trace.
Therefore, we can make predictions of whether the execution time would improve
if manual task annotations were actually translated into OpenMP pragmas, or if
alternative schedulers were available.
\added[id=R2]{
 Such statements are optimistic:
 For the present studies, we rely on FCFS scheduling with a global task queue. 
 We ignore NUMA effects as
 well as the critial path analysis, and we also neglect further external
 factors such as bandwidth constraints or task activation latencies.
}

\emph{Otter Vis} finally translates both the traced and hypothetical
data into HTML reports containing runtime metrics, graphs and figures.
It allows us to compare the recorded execution pattern of a code to different
simulated, hypothetical traces.
Such postprocessed data can guide the parallelization of a code, but also
uncovers in hindsight unfortunate scheduling decisions from a real run.

\section{Execution patterns}
\label{section:patterns}


For our demonstrator runs, a standard 16-core AMD EPYC\trademark Processor model 7302 serves as testbed,
although we intentionally limit the number of available OpenMP
threads to four.
This helps us to highlight execution patterns of interest.
With a larger number of threads, the resulting data can become too
multifaceted, obscuring details.
In all of our experiments, we have validated that the tracing induces negligible runtime overhead of less than 2\%.
We can trust the tracing data.

\subsection{Task spawn guarantees}

%
%
Task creation in the OpenMP API via a \texttt{task} directive introduces
a task and also constitutes a Task Scheduling Point (TSP).
At this point, it is at the OpenMP implementation's discretion
either to execute the task immediately ``in situ''
(undeferred in the same thread) or to actually spawn it as a deferred task.
The latter sends the task to the task pool, from which it is picked up
later; potentially by another thread.
While programmers can push the behaviour
towards undeferred execution via \texttt{if} and \texttt{final} clauses,
they can not enforce a deferred task.

This freedom is intended by the OpenMP specification.
It allows an implementation to easily deal with large numbers of tasks
by switching between undeferred and deferred execution as needed, depending on
runtime conditions such as number of available threads, current load of the
system, etc.
It allows for a throttling of the task creation
\cite{Agathos:2012:Throttling,Duran:2008:TaskThrottling}.

%
%

\begin{figure}[htb]
  \begin{center}
    \includegraphics[width=\textwidth]{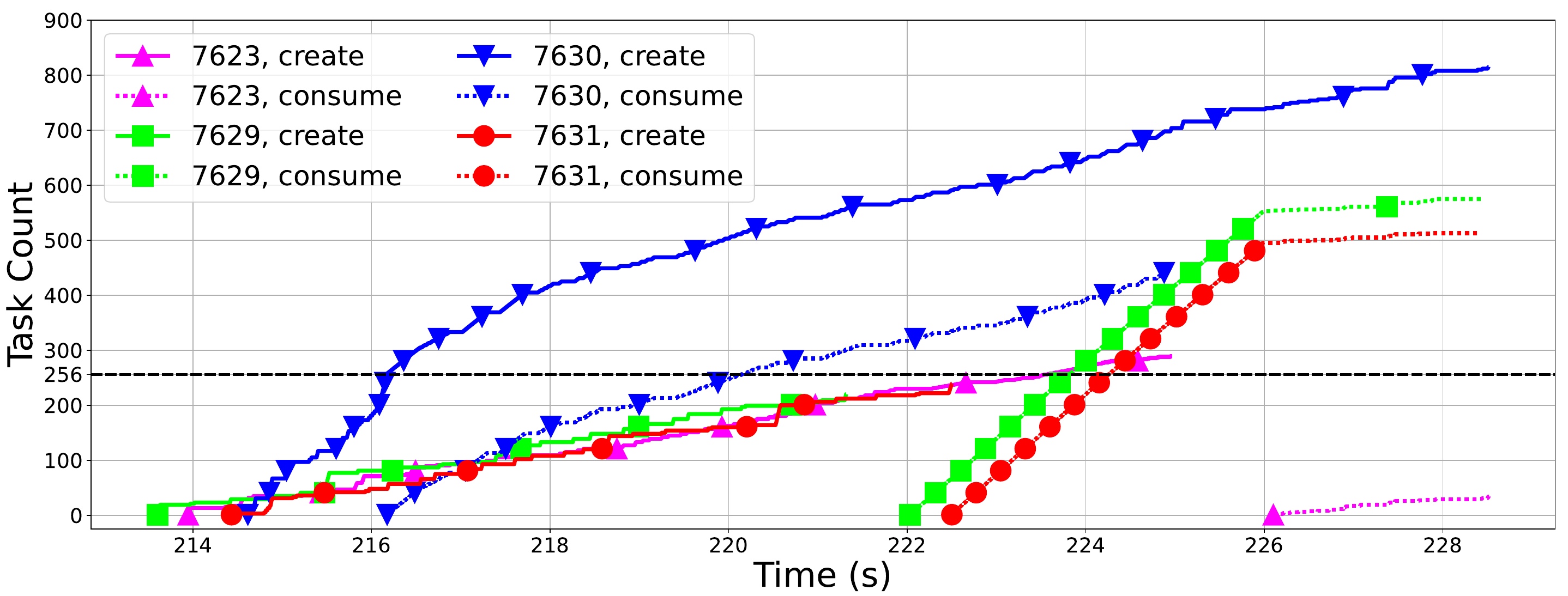}
  \end{center}
  \vspace{-0.5cm}
  \caption{
    \replaced[id=R1]{Created vs.~consumed \added[id=AT]{enclave} tasks on four threads (7623, 7629,
    7630, 7631) over time \added[id=AT]{for a single timestep}.
    7623, 7629 and 7631 produce tasks which are held in a task queue and then
    completed, i.e.~consumed once the producing task has terminated.
    7630 produces so many tasks that the
    producer task is suspended as further child tasks cannot be deferred. 
    They are executed immediately.
    From 222s onwards, 7629 and 7631 start to process tasks, eventually
    steal from 7630 and hence allow 7630 eventually to stop interrupting the
    producer. }{Left:
    Created vs.~consumed tasks on four threads over time.
    Right: Corresponding trace with grey bars illustrating the traversal threads
    (task producers), dots showing the creation of enclave child tasks, and
    vertical black bars denoting the actual enclave task execution. For
    immediate execution, the vertical bars are embedded directly into the
    producing traversal task. All data are recorded, i.e., not simulated.}
    \label{figure:task-recommendation:histogram}
  }
\end{figure}

As an implementation example, LLVM's OpenMP runtime maintains a double-ended
task queue per thread.
Each thread enqueues created tasks in its own task queue,
and always tries to pick a task for execution from the end.  
This leads
to an effective last-in first-out execution behaviour of tasks.
If there are no more ready tasks left in the thread-local queue, the thread attempts 
to steal tasks from queues of other threads.  Stolen tasks are taken from the
beginning of a queue to retrieve the longest-waiting tasks first.

\begin{figure}[htb]
  \begin{center}
    \includegraphics[width=\textwidth]{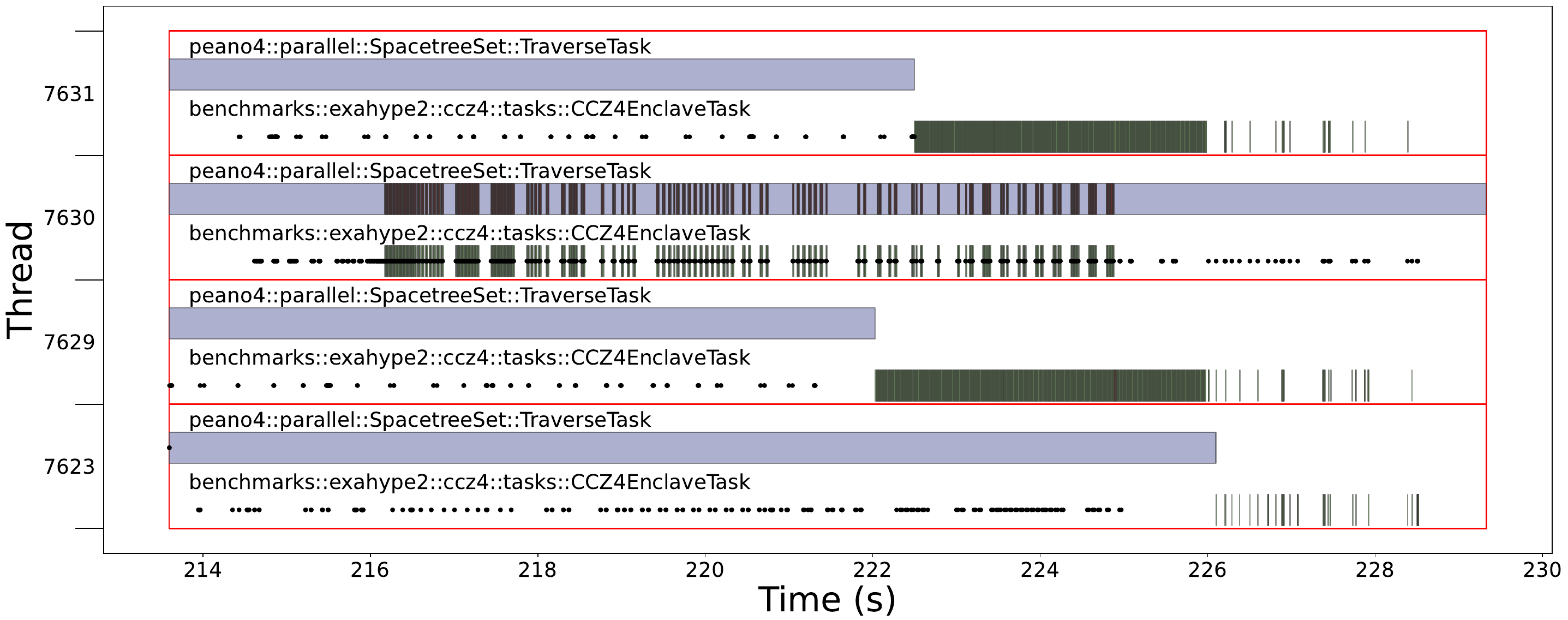}
  \end{center}
  \vspace{-0.5cm}
  \caption{  
    \replaced[id=R1]{Trace for task execution pattern from
    Figure~\ref{figure:task-recommendation:histogram} with grey bars illustrating the traversal \replaced[id=AT]{tasks}{threads (task producers)}, dots showing the creation of enclave child tasks, and
    \replaced[id=AT]{narrow}{vertical} black bars denoting the actual enclave task execution. \replaced[id=AT]{Bars embedded in the traversal task on thread 7630 show the task being suspended when the thread immediately executes an enclave task.}{For
    immediate execution, the vertical bars are embedded directly into the
    producing traversal task.} All data are recorded, i.e., not simulated.
    The trace illustrates that \replaced[id=AT]{the traversal task of thread 7630}{7630} is on the critical path.
    Not deferring child tasks prolongs this path.
    }{
    Left: Created vs.~consumed tasks on four threads over time.
    Right: Corresponding trace with grey bars illustrating the traversal threads
    (task producers), dots showing the creation of enclave child tasks, and
    vertical black bars denoting the actual enclave task execution. For
    immediate execution, the vertical bars are embedded directly into the
    producing traversal task. All data are recorded, i.e., not simulated.}
    \label{figure:task-recommendation:timeline}
  }
\end{figure}

%
%
In ExaHyPE, a traversal task spawns bursts of enclave tasks (Alg. \ref{algorithm:travesal}). Switching to undeferred mode implies that these enclave task bursts might be artificially constrained.
Traversal tasks terminating early due to geometric load imbalances consume
the tasks they have spawned before they continue to steal enclaves from
other threads' queues.
Traversal tasks running longer and producing many tasks see their tasks being
stolen by otherwise idle threads. 
To enable this behaviour is the intention and motivation behind our enclave
design~\cite{Charrier:2020:EnclaveTasking}.
Traversal tasks spawning a very high number of tasks might run
into a situation where they exceed their task queue size, while 
no other threads are available to steal their
tasks.
They stop further task production, and instead immediately process
child tasks~\cite{Schulz:2021:task_inefficiency_patterns}. 
Our testbed software stack seems to employ a queue threshold of 256 tasks.
Once a thread has enqueued more than 256 tasks, the system switches into the
undeferred mode (Figure~\ref{figure:task-recommendation:histogram}).

%
%
If tasks producing the lion's share of enclave tasks are also the critical tasks
within their fork-join section, the switch to an immediate consumption
introduces a bottleneck, as a failure to defer these tasks prolongs the
critical path (Figure~\ref{figure:task-recommendation:timeline}).
With \textit{Otter Sim}, we can simulate a world where the task
queues have no upper threshold, i.e., tasks are always deferred.
This would reduce the runtime by
up to 4.7~\%.
There are enough threads available towards the end of each traversal
to consume all deferred tasks spawned by the critical path.


%
%
There are two workarounds to enforce this behaviour: first,
we can introduce helper queues on top of the actual
OpenMP runtime \cite{Li:2022:TaskFusion,Schulz:2021:task_inefficiency_patterns}.
Rather than spawning OpenMP tasks directly, we hold
them back in a user-defined queue, releasing them only after the
production task has terminated.
While there are reasons for this approach besides the manual
deferring---it allows us to fuse withheld tasks into one meta task to deploy
them en bloc to a GPU \cite{Wille:2023:GPUOffloading} or to vectorize
aggressively \cite{Li:2022:TaskFusion}---it replicates OpenMP
core functionality.
The alternative second workaround \replaced[id=F]{instructs OpenMP to increase
its task queues upon demand (\texttt{KMP\_ENABLE\_TASK\_THROTTLING=0}). Yet,}{is
to increase the task queue sizes globally. While OpenMP implementations allow
for this, } there are good reasons for limited queue sizes. 
If we increase them dramatically\replaced[id=F]{, }{---irrespective that the
maximum size often is not known in dynamic AMR codes---}we have to pay a runtime and
memory overhead penalty\deleted[id=F]{ globally}.
\added[id=F]{
 Our demonstrator selectively identifies tasks which benefit from flexible queue
 sizes without asking for globally dynamic queues. 
}

%
%
%

\subsection{Nested parallelism}

%
%
High performance computing codes tend to combine task and data parallelism.
Modern codes also tend to rely on hierarchical parallelism
\cite{Royuela:2019:cooperative_parallel,Sun:2020:nested_parallelism}.
\replaced[id=F]{Such}{Executing such} a code makes the concurrency fan out as
the code descends along the call tree.

In the OpenMP API, the maximum number of nested parallelization levels is controlled through
\texttt{OMP\_MAX\_ACTIVE\_LEVELS}~\cite{OARB21}.
\replaced[id=F]{By default}{More importantly}, OpenMP realizes strictly nested
parallelism, i.e., \replaced[id=R1]{our data-parallel region mapped onto a
\texttt{parallel for}}{a data-parallel region} can only use a subset of the
threads available to the enclosing section.
A task that 
contains a \texttt{\deleted[id=F]{parallel} for} directive will therefore only utilize one
thread to execute that \deleted[id=F]{parallel} region, as each task is tied to one thread.


\begin{figure}[htb]
 \begin{center}
  \includegraphics[width=\textwidth]{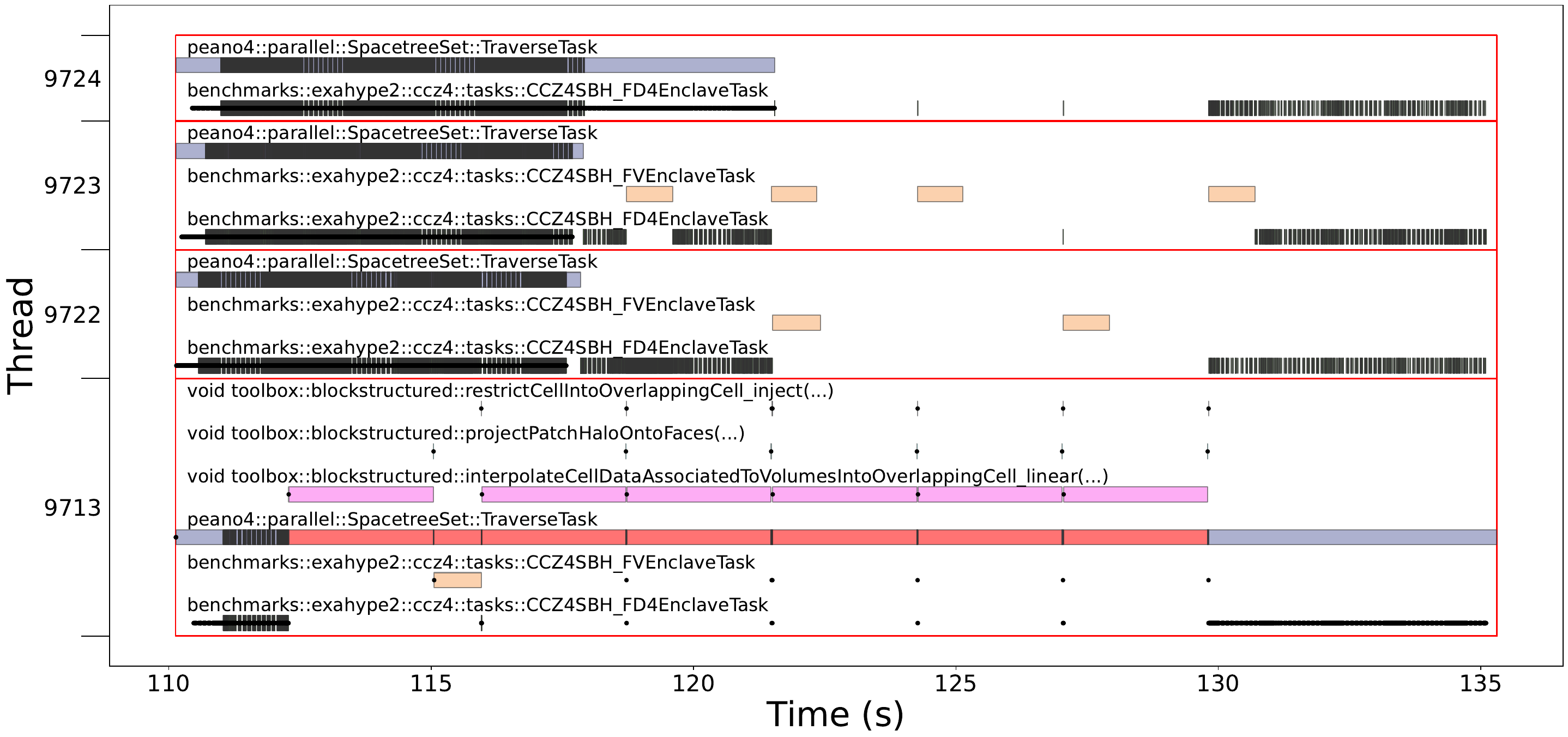}
 \end{center}
 \vspace{-0.4cm}
 \caption{
  \added[id=R1]{
  Timeline of a single time step. The red bars highlight where the
    critical task runs into some embedded \texttt{parallel for} constructs.
    Recorded timings augmented by postprocessing data (critical path).}
   \label{figure:nested-parallelism-timeline}
 }
\end{figure}

%
%
In ExaHyPE, the subdomain traversals are realized as tasks. 
The domain decomposition geometrically makes some tasks responsible for the
interpolation and restriction along AMR boundaries and/or the coupling of one
physical model to the other.
Both types of operations are very expensive.
Logically, the arising projections between different solvers or mesh
resolutions are embarrassingly parallel, i.e., could be mapped onto an embedded
\texttt{parallel for}.
Yet, we serialize any \texttt{parallel for} embedded into a
task.

%
%
As the loops are expensive and as they often align along the critical path---in
many ExaHyPE settings they are indeed responsible for making the owning task
a member of the critical path---the lack of support for nested parallelism
within tasks increases the makespan of the application (Fig.~\ref{figure:nested-parallelism-timeline}).
For characteristic demonstrator setups, we could, in theory, gain up to a 43~\%
reduction in runtime if the loops were executed in parallel.
This optimistic estimate assumes that sufficient idling threads are available
whenever a critical task encounters a loop.

%
%
The OpenMP API provides no mechanism that allows a task to
\replaced[id=F]{book}{execute on} multiple threads in a data-parallel fashion.
This leaves us with two alternatives.
On the one hand, we could \replaced[id=F]{use}{embed} a \texttt{parallel for},
i.e., fire up a new team of threads\deleted[id=F]{, into a task}.
On the other hand, we could switch from a \texttt{for} loop to a
\texttt{taskloop}.
In an HPC context, a new team of threads is problematic.
Typically, HPC codes spawn one thread per core right from the start to avoid
oversubscription\added[id=F]{, i.e.~hyperthreading} resulting in overheads due
to thread swapping.
\replaced[id=F]{
 Mechanisms to reuse existing threads within nested regions are available
 (cmp.~\texttt{KMP\_HOT\_TEAMS\_MAX\_LEVEL}), but do not address the present
 issue, if the threads are already booked out.
}{
Plus, applications do not want to pay the price for additional thread creation.}
Using a \texttt{taskloop} construct is hence more natural and, in combination
with a \texttt{default(shared)} clause, a minimally invasive change to the code.
Yet, creating \replaced[id=R1]{many subtasks through a}{a (potentially huge)
number of iteration tasks for a} \texttt{taskloop} introduces overhead in itself, while the tasks might
end up in the thread-local queue, i.e., continue to be serialized.

\subsection{Fair yields}

%
%
In ExaHyPE, there are typically few traversal tasks compared to the available core
count.
In our experiments, the traversal tasks therefore almost never wait
for the enclave\replaced[id=F]{ tasks}{s}. 
All enclave tasks are completed by threads not involved in the traversal ahead
of time.
There are exceptions to the rule: 
NUMA-intensive systems such as the AMD EPYC\trademark processors make some developers deploy one MPI process
per NUMA domain---effectively yielding a low core count per process---and dynamic
load balancing might deploy more subpartitions to an MPI process than there are
threads available.
Our implementation therefore protects the access to enclave task outcomes with
an atomic flag signalling \replaced[id=R1]{whether}{weather} the task is
complete.
If unset, the consumer thread yields and then polls again.

%
%
The result code starves in rare situations.
Let $T$ be the number of threads available and let $C>T+1$ be the
number of consumer \replaced[id=R1]{tasks}{threads} that traverse the domain and
consume one of the $E$ enclave task outcomes which are pending in the system.
$E+C \gg T$ then is the total number of tasks.
ExaHyPE can run into a situation where a consumer task yields, another untied
consumer is swapped in, and the consumers all take turns checking the enclave
task flags.
The enclave tasks \replaced[id=F]{then starve}{are starving}.

%
%
In the OpenMP API specification, the \texttt{taskyield} construct is a hint to
the implementation to introduce an additional TSP and, hence, to give other
tasks a chance to be scheduled for execution.  As a hint, an
implementation can ignore the TSP\replaced[id=R2]{ or implement it in various
ways with different performance implications \cite{Schuchart:2018:taskyield_and_mpi}.
For ExaHyPE, it is problematic that }{.
While ignoring is conforming from the perspective of the OpenMP specification,
it is not ideal for ExaHyPE and other projects, such as \cite{Schuchart:2018:taskyield_and_mpi}.

%
%
}OpenMP's \texttt{taskyield} does not provide a fairness guarantee and notably
cannot be used to drain a pending task queue incrementally.
One might argue that a \replaced[id=R4]{robust}{proper} realization of our
producer-consumer pattern should employ task dependencies to
\deleted[id=R4]{induce a proper task execution order and to} avoid the
\deleted[id=R4]{explicit} polling for task readiness.
Yet, we note that \texttt{taskwait}
with a \texttt{depend} clause is \deleted[id=R4]{a fairly new construct and} not
always straightforward to use from a programmer's point of view, as all task
dependency addresses have to reside within the user space.
More severely, task dependencies are restricted to sibling tasks in
OpenMP\added[id=R4]{ whereas we have parent--child relations here}.
ExaHyPE would benefit from a fair yielding mechanism\added[id=R3]{ to avoid
occasional deadlocks on some hardware. At the moment, we have to manually work
around such cases by adding user-defined task queues}.

\subsection{Taskwait semantics}

%
%
The \texttt{taskwait} and \texttt{taskloop} directives used by our traversal
tasks introduce synchronization points.
We use \texttt{taskwait} with \replaced[id=F]{\texttt{taskloop nogroup}}{the
\texttt{nogroup} clause} to synchronize the set of all child tasks with the
same parent task, i.e., all traversal siblings, but not their spawned enclave tasks, while the synchronization set for the \texttt{taskloop} without \texttt{nogroup}
includes all descendant tasks created from within the \texttt{taskloop}
region, i.e., all enclave tasks produced, too.
\added[id=F]{
 The \texttt{taskloop nogroup} can be replaced by a for loop spawning the
 traversal tasks individually, which is necessary for the NVIDIA software stack
 that lacks support for task groups.
 If supported, we find it to be more elegant and
 \replaced[id=R1]{slightly}{slighlty} faster than manual task spawning.
}

It is up to the  OpenMP implementation how the synchronization points are realized.
There are two basic options.
First, the runtime can process further tasks while waiting until all tasks in the
synchronization set have completed execution.  
Second, it can actively poll the synchronization construct, possibly deciding not to execute further tasks while it waits.

Let ExaHyPE spawn $K$ (traversal) tasks at one point, and immediately
after the end of that task group issue another one with $K$ tasks.
These are two time steps. If $K-1$ threads decide to process further (enclave) tasks at the first
synchronization point, the final remaining thread might hit the end of the
first task group while these threads still are busy.
Only this one thread hence is available to immediately continue with the 
$K$ tasks from the second task group.

%
%
While it makes sense for a thread to execute further
tasks while waiting in a \texttt{taskwait} or at the end of a \texttt{taskgroup}
region---this guarantees progress---it means that 
we add algorithmic latency to the second task group in the example.
This latency is defined by the 
time it takes the $K-1$ threads to finish the currently active enclave task,
and to join traversal tasks of the next task group, i.e., time step.


%
%
In ExaHyPE, the traversal tasks \deleted[id=R1]{have to have high priority.
They} define the critical path.
If one of the traversal tasks is not immediately kicked off at the start of a
task group, we \replaced[id=F]{risk delaying}{run risk that this delay extends into} the critical path.
Furthermore, we observe that such latency can lead to low occupancy further down
the road, as tasks have been processed at a scheduling point where it would have
been better from a performance standpoint if the underlying thread had
\replaced[id=R4]{paused for a moment}{waited briefly} and then
continued with the traversal task~\cite{Schulz:2021:task_inefficiency_patterns}.
Yet, OpenMP provides no mechanism to stop a thread at the end of a task group
from continuing with other tasks.
\added[id=R2]{
 We have no mechanism to flag to the system that we are aware that there are
 many ready tasks but that we also know that there will be a point reached soon
 with a low concurrency level where these tasks can all be handled without
 delaying any other time-critical task.
 In such a case, we might be willing to accept low occupancy temporarily, as
 long as we can trade this to an immediate continuation along the critical
 path---knowing that there will be enough resources later on to handle
 all the postponed tasks.
}

\section{OpenMP extensions and their realization}
\label{section:realisation}

\begin{algorithm}[htb]
  \scriptsize
  \algrenewcommand\algorithmicindent{0.4em}%

\begin{algorithmic}
 \Function{traversal}{\ldots}
 \State \#pragma omp ... nogroup latency
 \For{(int subdomain = 0; \ldots)}
   \State \ldots
   \While{\ldots}
     \State \#pragma omp taskyield throughput
   \EndWhile
   \State \ldots
   \State \#pragma omp task defer
   \State \ldots
 \EndFor
 \EndFunction
\end{algorithmic}

 \caption{
   \added[id=R1]{
    Domain traversal loop with modified OpenMP annotations.   }
    \label{algorithm:modified-traversal}
  }
\end{algorithm}

With a clear description of runtime flaws, we can propose some
modifications to the OpenMP API that would help our demonstrator code.
We distinguish between proposals for the specification
API and suggestions how to implement an altered specification
(Algorithm~\ref{algorithm:modified-traversal}).

\subsection{API modifications}

%
%
ExaHyPE with its producer-consumer pattern would benefit from
explicitly labelling tasks as ``must be deferred'',
This would complement the existing semantics of the OpenMP clauses \texttt{if}
and \texttt{final}, i.e., allow developers to disable task
throttling~\cite{Agathos:2012:Throttling,Duran:2008:TaskThrottling,Gautier:2018:task_granularity}.
In ExaHyPE, our first and foremost goal is the reduction of the critical path
requiring such a flag to be prescriptive.
Yet, realization constraints and task pool overflows might require it to become
a weakly prescriptive annotation (see below).  A possible extension of the OpenMP API
would be a new clause \texttt{defer} that extends the existing clauses of the \texttt{task}
directive: \texttt{\#pragma omp task defer(}\textit{condition}\texttt{)}.
If \textit{condition} evaluates to \textit{true}, the task shall be deferred; 
if it evaluates to \textit{false}, the task maybe undeferred or deferred.

ExaHyPE would benefit from the introduction of \deleted[id=R1]{a parallel} tasks 
that roll out embedded loops over multiple threads.
Providing such a feature with spreading guarantees is difficult
\cite{Sun:2020:nested_parallelism}\replaced[id=R1]{. A}{, but a} plain 
\texttt{taskloop} with   
\replaced[id=R1]{the clause \linebreak
\texttt{priority(omp\_get\_max\_task\_priority())}}{new
clauses \texttt{priority(max) grainsize(idlethreads)}} would
\added[id=R1]{not} facilitate the feature.
It would assign the resulting loop chunks high priority compared to any other
task in the system and \replaced[id=R1]{load}{split up the loop into as many
chunks as there are idle threads at this point.
Load} stealing would implicitly
scatter the iteration range among the available OpenMP threads.
\added[id=R1]{However, there is no guarantee that they are stolen. We would required a \texttt{scatter}
clause.}

For a \texttt{taskyield} variant, we would envisage that programmers should
be able to decide that a \texttt{taskyield} region should not be an no-op,
but actually pick tasks from the task pool for execution.
Furthermore, it would help to mark a TSP of a \texttt{taskyield} region
as either high throughput or low latency: \texttt{\#pragma omp taskyield latency\textbar throughput},
with the default being the current implementation-defined behaviour.

A low latency TSP suspends the encountering task yet brings it back as soon as possible to
minimize the probability that we lose all of its cache content, following the depth-first philosophy of OpenMP implementations.
It reduces the algorithmic latency of polling realized through yield.
In our case, we would rather use a throughput-oriented TSP which would cause a yielding task to go to the back of a queue.
Such a yield could then also provide fairness guarantees.

To facilitate low latency \texttt{taskwait} or \texttt{taskgroup} regions, the above \texttt{latency} and \texttt{throughput} clauses would also be added to these directives.
It instructs the implementation that \replaced[id=R1]{threads hitting the
corresponding synchronization point}{the used threads} are to be kept free of
other tasks, as they will be needed immediately afterwards for some high priority work (\texttt{latency}).
While a suspended task is waiting for tasks to complete, the implementation shall not schedule other tasks
to reduce wakeup latency.  It is up to the user to guarantee that this clause does not induce a
deadlock or starvation, for example by having all tasks encounter a latency optimized \texttt{taskwait latency} directive.

\subsection{Realization}

The opportunity to manually defer tasks to the task pool independent of the execution context
means that we have to provide a dynamic task queue which can grow without any
constraints as a realization of an unbounded task pool.
Otherwise, task pools might overflow.
A weakened realization sticks to the existing implementation of task queues, but instead switches
from task stealing to task distribution for ``must be deferred'' tasks:
Whenever a thread spawns more tasks than its local task queue is able to
accommodate, these tasks first are scattered over other threads' queues (similar
to how tasks with data affinity are distributed~\cite{Klinkenberg:2018:data_affinity}).
If and only if this task distribution fails as well, we fall back to undeferred
execution for the created task.
We hypothesize that this last variant provides a reasonable compromise between
runtime efficiency, small changes to existing infrastructure, and improved
runtime characteristics on the demonstrator side.
To avoid that an active task distribution confuses the scheduling of victim
threads, it is important that the created subtasks are assigned a very low
priority, i.e., are not brought forward on the target thread. 
Otherwise, we would replicate the motivating problem once again.

To allow parallel loops that are embedded into a task to ``invade'' other
threads, an implementation should go the other way regarding task priorities.
The loop segments scattered over the queues have to have
a higher priority than the highest priority task in any respective queue, and 
task stealing has to take priorities into account, too.
This way, we can ensure that idle tasks steal the ``right'' tasks from the
thread issuing the parallel for, and that the stealing does not delay
the actual execution of the parallel loop that kicked it off.

A fair yield which can guarantee progress in our particular case could easily be
mapped onto priorities, too:
If a thread queue is aware of the lowest priority task, a throughput yield would
label the suspended task with a priority that is by one smaller than the
currently lowest priority.

The new ``low latency'' clause finally requires us to eliminate the task
scheduling point at the end of the loop.
This will let the used threads (logically) run idle.
Once we ensure that a subsequent parallel loop schedules ``its own'' tasks or
parallel segments first, we however obtain a taskwait which
prioritizes low algorithmic latency over throughput.
To ensure that the subsequent loop start does not issue any other task first, we
can either hardcode the scheduling or make the scheduling biased.
Given the spawned task higher priorities than all other pending ready tasks
introduces the required bias.

\subsection{Contextualization}

There are many valid reasons and rationale why OpenMP implementations fall back to task 
throttling if too many tasks are created~\cite{Gautier:2018:task_granularity}.
Our suggestion to introduce a dedicated new clause to avoid this accepts this
fact, as it suggests localized modifications to few tasks.
Such a clause should have no detrimental effect on existing codes.

Our nested parallelism within tasks aligns with existing developments within
OpenMP and does not sacrifice threads \cite{Sun:2020:nested_parallelism}.
It notably fits to OpenMP's GPU kernel concept, where the \texttt{target} pragma
lets the spawning code fan out into a new team with many threads.
With a full support of nested parallelism on the host, massive tasks that should be offloaded to a GPU yet cannot be moved there for
one reason or the other benefit from the full concurrency of the host processor.
The feature facilitates platform- and performance-portable code.

\added[id=F]{
 A fair yield is a natural cousin to the existing \texttt{detach} clause, which
 becomes useful if we cannot easily construct or realize a completion check and
 instead prefer (partial) draining of the task queue.
}

A ``do not schedule'' policy at an implicit synchronization point stands in the
tradition of OpenMP to grant NUMA considerations high priority.
The probability is high that people use it for subsequent parallel loops with
the same granularity which run over related data.

\section{Conclusion and outlook}
\label{section:conclusion}

OpenMP\replaced[id=F]{ schedules are}{'s scheduling conventions are} often not
\added[id=F]{unique or} enforced by the standard\deleted[id=F]{, and some are
labelled as optional}.\deleted[id=F]{ As all mainstream implementations
commit to similar implementations (such as depth-first scheduling, ignoring priorities, and stealing the oldest tasks first), developers may assume certain implementation-defined scheduling patterns are part of the standard.}
We introduce four scenarios, where \replaced[id=R1]{the LLVM
implementation introduces}{mainstream scheduler implementations introduce}
runtime flaws:
tasks are prematurely activated,
tasks do not support embedded parallelism,
tasks do not yield in a fair way,
and waiting constructs always prioritize high throughput instead of low algorithmic
latency.

\added[id=F]{
 These flaws result from common interpretations and
 rationale how to interpret and realise the standard efficiently for a magnitude of applications.
}
Our work does not challenge the underlying implementation rationale of
mainstream runtimes---indeed there are good reasons to implement things the way
they are---but it suggests that users should be allowed to explicitly instruct OpenMP to realize things
differently.
While prescriptive OpenMP statements already enforce certain OpenMP behaviour,
our work goes one step further and makes the prescriptive character cover
certain realization decisions, too.


These modifications do not require major rewrites of the OpenMP runtime.
Instead, the majority of them can be implemented
\replaced[id=F]{combining}{through proper support of} task
priorities\replaced[id=F]{ with minor changes in the runtime's logic.
\replaced[id=R1]{For}{or} all changes}{. We deduce that} a mature
implementations of priorities is a sine qua non\replaced[id=F]{ which can
induce further scalability challenges on massively parallel systems}{ for future
OpenMP runtimes}.
In combination with few further tweaks, they \added[id=F]{however} will provide
multifaceted tuning opportunities to codes like ExaHyPE.

\added[id=R1]{
 GNU and other runtimes share implementation rationale with LLVM. 
 We may therefore expect that many of the documented
 flaws arise there, too, likely with quantitatively different
 characteristics.
 It is future work to assess these differences systematically.
 Our work uses one bespoke simulation code as demonstrator.
 We again expect other task-heavy codes to encounter similar flaws and, hence,
 to benefit from the proposed extensions.
 The scientific challenge for future work is to quantify these effects, but also
 to identify if the extensions and required modifications of the runtime could
 potentially harm the performance of other codes.
}
\added[id=R2]{
 They have the potential to make runtime implementations not backward compatible
 from a performance point of view.
}

\section*{Acknowledgments}

Tobias' research has been supported by EPSRC's Excalibur
programme through its cross-cutting project EX20-9 \textit{Exposing Parallelism: Task Parallelism}
(Grant ESA 10 CDEL) and the DDWG projects \textit{PAX--HPC} (Gant EP/W026775/1)
as well as \textit{An ExCALIBUR Multigrid Solver Toolbox for ExaHyPE}
(EP/X019497/1).
His group appreciates the support by Intel's Academic Centre of
Excellence at Durham University. 
The comparison of OpenMP vs.~TBB and the assessment of early oneAPI OpenMP
behaviour has led to some of the investigations reported here.
This work has made use of the Hamilton HPC Service of Durham University.

AMD, the AMD Arrow logo, EPYC, and combinations thereof are trademarks
of Advanced Micro Devices, Inc. Other product names used in this publication are for identification purposes
only and may be trademarks of their respective companies.
ExaHyPE \footnote{\fontsize{8}{8}\selectfont
\url{https://gitlab.lrz.de/hpcsoftware/Peano/-/releases/2024OpenMPPaper}} and
Otter \footnote{\fontsize{8}{8}\selectfont
\url{https://github.com/Otter-Taskification/otter/releases/tag/2024-openmp-paper}}
\footnote{\fontsize{8}{8}\selectfont
\url{https://github.com/Otter-Taskification/pyotter/releases/tag/2024-openmp-paper}}
are open source.

\appendix

\bibliographystyle{plain}
\bibliography{paper}

\end{document}